\documentclass[12pt,draftcls,onecolumn,twoside]{IEEEtran}
\usepackage[T1]{fontenc}
\usepackage[utf8]{inputenc}
\usepackage{color}
\usepackage{pgfplots} 
\pgfplotsset{compat=newest} 
\pgfplotsset{plot coordinates/math parser=false} 
\usepackage[noadjust]{cite}
\usepackage[cmex10]{amsmath}
\usepackage{amsfonts}
\usepackage{amssymb}
\usepackage{ntheorem}
\usepackage{dsfont}    
\usepackage{mathtools} 
\DeclareMathOperator{\sign}{sgn}
\usepackage{esvect}
\usetikzlibrary{arrows}
\usepackage[tight,footnotesize]{subfigure}
\theoremstyle{plain}
\theoremheaderfont{\itshape}\theorembodyfont{\itshape}
\theoremseparator{.}

\newtheorem{proposition}{Proposition}

\theoremheaderfont{\itshape}\theorembodyfont{\upshape}
\theoremseparator{.}

\newtheorem{remark}{Remark}

\begin{document}
%
%
%
\title{Energy-Efficient Classification for Anomaly Detection: The Wireless Channel as a Helper}%
%
\author{Kiril~Ralinovski, Mario~Goldenbaum, and S\l awomir~Sta\'{n}czak%
\thanks{K.~Ralinovsiki and S.~Sta\'{n}czak are with the Network Information Theory Group, Technische Universität Berlin, 10587 Berlin, Germany. M.~Goldenbaum is with the Department of Electrical Engineering, Princeton University, Princeton, NJ 08544, USA.}}
%
\maketitle
\vspace{-25pt}
\begin{abstract}
Anomaly detection has various applications including condition monitoring and fault diagnosis. The objective is to sense the environment, learn the normal system state, and then periodically classify whether the instantaneous state deviates from the normal one or not. A flexible and cost-effective way of monitoring a system state is to use a wireless sensor network. In the traditional approach, the sensors encode their observations and transmit them to a fusion center by means of some interference avoiding channel access method. The fusion center then decodes all the data and classifies the corresponding system state. As this approach can be highly inefficient in terms of energy consumption, in this paper we propose a transmission scheme that exploits interference for carrying out the anomaly detection directly in the air. In other words, the wireless channel helps the fusion center to retrieve the sought classification outcome immediately from the channel output. To achieve this, the chosen learning model is linear support vector machines. After discussing the proposed scheme and proving its reliability, we present numerical examples demonstrating that the scheme reduces the energy consumption for anomaly detection by up to 53\,\% compared to a strategy that uses time division multiple-access.
\end{abstract}
%
%
%
\section{Introduction}
Classification for anomaly detection is a common objective in areas such as image recognition, condition monitoring, and fault diagnosis \cite{Widodo07}. The corresponding procedure is typically divided into two phases. During the first phase, the system environment is observed over a certain period of time in order to learn what the normal system state is. In the second phase, the instantaneous system state is sensed and then classified according to some rule (i.e., classifier) derived from the first phase. If the learning model was appropriately chosen, the classifier output indicates whether the system is currently in normal or abnormal state, where the latter is what is called an \emph{anomaly}  \cite{Schoelkopf99}. In industrial automation, for instance, the system state (or system condition) is monitored by means of spatially distributed temperature, oil pressure, and vibration sensors, just to name a few, that have a wired connection to some central processing unit. 
\begin{figure}
	\centering
	\begin{tikzpicture} 
	\node[draw](A) at (0.2,-0.4)[minimum width=2.3cm]{\small Sensor Node $1$};
	\draw (A) -- (1.6,-0.4);
	\draw (1.6,-0.4) -- (1.6,0);
	\draw (1.6,0) -- (1.4,0.2);
	\draw (1.6,0) -- (1.8,0.2);
	\node[draw](B) at (0.2,-2.4)[minimum width=2.3cm]{\small Sensor Node $K$};
	\draw (B) -- (1.6,-2.4);
	\draw (1.6,-2.4) -- (1.6,-2.0);
	\draw (1.6,-2.0) -- (1.4,-1.8);
	\draw (1.6,-2.0) -- (1.8,-1.8);
	\draw (1.6,-1) node {$\vdots$}; 
	\draw[thick,->] (2.0,0) -- (4.0,-0.8);
	\draw[thick,->] (2.0,-2) -- (4.0,-1.2);
	\draw[thick,->] (2.0,-1) -- (4.0,-1);
	\node[draw](C) at (5.7,-1.4)[minimum width=2.3cm]{\small Fusion Center};
	\draw (4.3,-1) -- (4.1,-0.8);
	\draw (4.3,-1) -- (4.5,-0.8);
	\draw (4.3,-1) -- (4.3,-1.4);
	\draw (C) -- (4.3,-1.4);
	\draw [thick, draw=black, fill=white]
	(4.5,-4) -- (7.0,-4) -- (7.0,-3.4) -- (4.8,-2) -- (4.5,-2) -- cycle;
	\draw [thick, draw=black, fill={rgb:black,1;white,3},]
	(7.0,-3.4) -- (7.0,-2) -- (4.8,-2) -- cycle;
	\fill[black] (4.77,-2.93) circle (0.1cm);
	\draw (5.53,-3.8) node {\small Normal State}; 
	\draw (6.29,-2.22) node {\small Anomaly}; 
	\path[->] (6.0,-1.7)  edge [bend right] (4.8,-2.8) ;   
	\end{tikzpicture}
	\caption{In (wireless) anomaly detection, the objective is to classify an instantaneous system state based on the observations taken by a set of distributed sensor nodes. A conventional transmission strategy would provide that all the raw measurements are transferred to a fusion center where the classification is carried out.}
	\label{fig:task}
\end{figure}
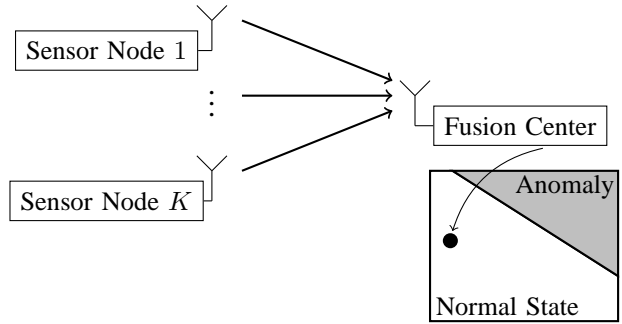

A more flexible and cost-effective way to deal with it is to use a wireless sensor network (WSN) \cite{Willig:08,Liqun12}. A WSN, however, requires sensor nodes to be driven by batteries whose replacement is generally too expensive. Energy efficiency is therefore a major design objective in order to guarantee sufficiently long network lifetimes \cite{Rault14,Akyildiz02}. A significant step in this direction is to drastically reduce the energy needed for communication. In a traditional WSN, the communication is as follows (see Fig.~\ref{fig:task}): the sensor nodes encode their observations and transmit them to a designated fusion center (FC) by means of some interference avoiding channel access method such as time division multiple-access (TDMA) or carrier sense multiple-access (CSMA). The FC then decodes all the data and classifies the instantaneous system state. 

As this strategy can be highly inefficient in terms of energy consumption, in this paper we propose an anomaly detection scheme that is based on the communication strategy proposed and evaluated in \cite{Goldenbaum:Stanczak:13a,Kortke:Goldenbaum:Stanczak:14,Goldenbaum:Stanczak:Boche:15}. The big difference to the traditional strategy outlined above is that \emph{interference is exploited}, rather than avoided, for carrying out the anomaly detection directly \emph{in the air}. In other words, the wireless channel helps the FC to retrieve the sought classification outcome immediately from the channel output. To achieve this, the learning model chosen in this paper is \emph{linear support vector machines} \cite[p. 112]{Cristianini00}. Due to its universality it is a widely used model and has recently been proven to be also a good candidate for condition monitoring and fault diagnosis \cite{Widodo07}. After adapting the transmission scheme of \cite{Goldenbaum:Stanczak:13a} to our specific needs, we prove that the resulting classification scheme is able to detect anomalies over the wireless channel with arbitrary small probability of error. In this context, we also discuss the trade-off between achievable accuracy and communication costs in terms of the number of transmissions needed. Finally, we present some numerical examples that demonstrate that the corresponding energy consumption can be reduced by up to 53\,\% compared to a strategy that uses TDMA.
%
%
%
\subsection{Paper Organization} \label{sec:organization}
The rest of the paper is organized as follows. Section~\ref{sec:system_model} provides the system model and the problem statement. In Section~\ref{sec:support}, we summarize the basics of linear support vector machines as the chosen learning model. Subsequently, in Section~\ref{sec:wireless_helper} we propose a transmission scheme that exploits interference for classifying system states and prove its performance. Section~\ref{sec:numerical_examples} provides some numerical examples and comparisons with a standard schemes, whereas Section~\ref{sec:conclusions} concludes the paper.

%
%
%
\subsection{Notational Remarks} \label{sec:notation} 
Random variables are denoted by uppercase letters and their realizations by lowercase letters, respectively. Vectors are denoted by bold lowercase letters and matrices by bold uppercase letters. The Euclidean norm of some vector $\mathbf{x}$ is denoted as $\|\mathbf{x}\|$. The expected value and the variance of some random element are denoted as $\mathbb{E}\{\cdot\}$ and $\mathbb{V}\mathrm{ar}\{\cdot\}$, whereas $\mathbb{P}(A)$ designates the probability of any event $A$. Finally, $\mathcal{N}(\boldsymbol{\mu},\mathbf{\Sigma})$ represents the multivariate normal distribution with mean vector $\boldsymbol{\mu}$ and covariance matrix $\mathbf{\Sigma}$.
%
%
%
\section{System Model and Problem Statement} \label{sec:system_model}
Consider a WSN that consists of a FC and a finite number of $K \geq 2$ spatially distributed nodes. As depicted in Fig.~\ref{fig:task}, the objective of the network is to monitor some system and periodically sample and classify its state in order reliably and efficiently detect anomalies. Towards this end, each node observes a coordinate of the corresponding $K$-dimensional system state $\mathbf{s}[t]\coloneqq(S_1[t],\dots,S_K[t])^T$, $t\in\mathds{N}$, assumed to be an element of the state space $\mathcal{S}\coloneqq\mathcal{S}_1\times\dots\times\mathcal{S}_K$. Here and hereafter, $\mathcal{S}_k\subset\mathds{R}$, $k=1,\dots,K$, represents the compact interval of values sensor $k$ can provide. We simply call this the \emph{sensing range} of node $k$, which can be a range of temperatures, pressures, vibration intensities, etc. 

We model the sensors' observations $\{S_k[t]\in\mathcal{S}_k\}_{t\in\mathds{N}}$, $k=1,\dots,K$, as time-discrete stochastic processes.\footnote{This implies that system state $\mathbf{s}[t]$ is a time-discrete vector-valued stochastic process.} Each node is equipped with a transmitter in order to send its instantaneous observation to the FC. Thus, for every fixed $t$, each node, say node $k$, maps $S_k[t]$ to a length-$M$ sequence of complex-valued transmit symbols $X_k^{(t)}[1],\dots,X_k^{(t)}[M]$, subject to some peak-power constraint $P_{\mathrm{max}}>0$, that is, $\max_{1\leq m\leq M}|X_k^{(t)}[m]|^2\leq P_{\mathrm{max}}$. Allowing the nodes to transmit concurrently in the same frequency band, the sequence of symbols received by the FC can be modeled as \cite{ElGamal:Kim:11}
\begin{equation}
	Y^{(t)}[m] = \sum_{k=1}^K h_kX_k^{(t)}[m] + N^{(t)}[m]\;,\;m=1,\dots,M\;.
	\label{eq:wmac}
\end{equation}%
Coefficient $h_k\in\mathds{C}$ denotes the attenuation between node $k$ and the FC, assumed to be known to node $k$, and $N$ a proper complex Gaussian receiver
noise process of variance $\sigma_N^2$.

Assume that state space $\mathcal{S}$ can be divided into classes ``normal'' and ``anomaly'' by a hyperplane. This means that at every time instant $t$, the system state $\mathbf{s}[t]$ belongs to either of these two classes. For ease of notation, we describe this by the set $\mathcal{C}\coloneqq\{\pm 1\}$, where $-1$ represents the normal states and $+1$ anomalies. Then, the problem to be solved in this paper is to first learn from a subset of the sensors' observations a \emph{classifier} 
\begin{equation}
	f:\mathcal{S}\to\mathcal{C}\,,\,\mathbf{s}\mapsto f(\mathbf{s})
	\label{eq:classifier}
\end{equation}%
that is optimal in the sense that it minimizes false classifications. Given such classifier, we then wish to efficiently compute it over the wireless channel (\ref{eq:wmac}) such that 
\begin{equation}
	\forall\mathbf{s}\in\mathcal{S}:\lim_{M\to\infty}\mathbb{P}\bigl(f(\mathbf{s})\neq\hat{f}(\mathbf{s})\bigr)=0\;,
	\label{eq:prob_condition}
\end{equation}%
where $\hat{f}$ denotes some estimator of $f$.
%
%
%
\section{Linear Support Vector Machines} \label{sec:support}
Learning the classifier (\ref{eq:classifier}) usually requires to choose an appropriate model. Due to its universality and simplicity, we therefore consider in the following the supervised learning model \emph{linear support vector machines} \cite{Cristianini00}. To this end, we shall assume that we are provided with a \emph{training set}, which is a sequence of already classified system states. To be more precise, for some $L\in\mathds{N}$, the training set is defined as
\begin{equation}
	\mathcal{T}\coloneqq\left\{(\mathbf{s}[l],c_l)\,\big|\,\mathbf{s}[l]\in\mathcal{S},c_l\in\mathcal{C},l=1,\dots,L\right\}
	\label{eq:training_set}
\end{equation}%
without loss of generality. Thus, for some realization of the system state, $\mathbf{s}[l]$ denotes its $l$-th sample and $c_l$ the class to which it belongs (i.e., $-1$ or $1$). Given this training set, we are interested in determining an appropriate classifier $f$ that is then used to reliably classify the remaining samples $\mathbf{s}[L+1],\mathbf{s}[L+2],\dots$ of the system state.

For linear support vector machines, (\ref{eq:training_set}) is supposed to be separable by a hyperplane 
\begin{equation}
	\mathbf{w}^T\mathbf{s}+b=0\;,\;\mathbf{s}\in\mathcal{S}\;,
	\label{eq:hyper_plane}
\end{equation}%
determined by some nonvanishing normal vector $\mathbf{w}\in\mathds{R}^K$ and some offset $b\in\mathds{R}$. In this context, (\ref{eq:hyper_plane}) is said to be \emph{optimal} if it divides the training samples having $c_l=-1$ from those having $c_l=+1$ by a maximal margin \cite{Vapnik95}.\footnote{Note that maximizing the margin minimizes false classifications.} Furthermore, (\ref{eq:hyper_plane}) is said to be \emph{canonical}, with respect to $\mathcal{T}$, if $\mathbf{w}$ is normalized such that \cite{Vapnik95}
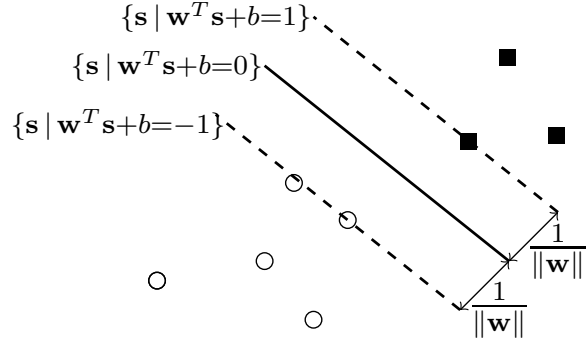
\begin{figure}
	\centering
	\scalebox{1.3}{
	\begin{tikzpicture} 
		\draw[<->] (5.5,0) -- (5.0,-0.5);
		\draw[<->] (5.5,0) -- (6.0,0.5);
		\draw[thick] (5.5,0) -- (3,2);
		\draw[thick,dashed] (5.0,-0.5) -- (2.5,1.5);
		\draw[thick,dashed] (6.0,0.5) -- (3.5,2.5);
		\draw (2.5,2.5) node {$\scriptstyle \{\mathbf{s}\,|\,\mathbf{w}^T\mathbf{s} + b = 1 \}$}; 
		\draw (1.5,1.4) node {$\scriptstyle \{\mathbf{s}\,|\,\mathbf{w}^T\mathbf{s} + b = -1 \}$}; 
		\draw (2.0,2.0) node  {$\scriptstyle \{\mathbf{s}\,|\,\mathbf{w}^T \mathbf{s} + b = 0 \}$}; 
		\draw (3.3,0.8) circle (0.085cm);
		\draw (3.85,0.42) circle (0.085cm);
		\draw (3.5,-0.6) circle (0.085cm);
		\draw (3.0,0) circle (0.085cm);
		\draw (1.9,-0.2) circle (0.085cm);
		\draw (1.9,-0.2) circle (0.085cm);
		\fill[black]  (5.0,1.14) rectangle (5.17,1.31);
		\fill[black] (5.9,1.2) rectangle (6.07,1.37);
		\fill[black]  (5.4,2.0) rectangle (5.57,2.17);
		\draw (5.4,-0.5) node {$\frac{1}{\| \mathbf{w} \| }$}; 
		\draw (6.0,0.1) node {$\frac{1}{\| \mathbf{w} \| }$}; 
	\end{tikzpicture}}
	\caption{A training set of system states that can be separated by a hyperplane. In this example, the \emph{optimal} separating hyperplane is represented by the solid line. The dotted lines represent the boundary of the margin. State samples that lie on the boundary are called support vectors.}
	\label{fig:optimal_hyperplane}
\end{figure}%
\begin{equation*}
	\label{eq:CanHyp}
	\min_{1\leq l\leq L}\bigl|\mathbf{w}^T\mathbf{s}[l]+b\bigr|=1\;.
\end{equation*}%
See Fig.~\ref{fig:optimal_hyperplane} for an illustration. Notice that if (\ref{eq:hyper_plane}) is canonical, none of the training samples $\mathbf{s}[1],\dots,\mathbf{s}[L]$ lies inside the margin bounded by the two hyperplanes $\mathbf{w}^T\mathbf{s}+b=\pm 1$, $\mathbf{s}\in\mathcal{S}$. That is, for $l=1,\dots,L$ 
\begin{IEEEeqnarray}{rCl}
	\mathbf{w}^T\mathbf{s}[l]+b\geq +1\quad&\text{if}&\quad c_l=+1\nonumber\\
	\mathbf{w}^T\mathbf{s}[l]+b\leq -1\quad&\text{if}&\quad c_l=-1\;,\nonumber
\end{IEEEeqnarray}%
which can be summarized to
\begin{equation}
	c_l\bigl(\mathbf{w}^T\mathbf{s}[l]+b\bigr)\geq 1\;,\;l=1,\dots,L\;.
	\label{eq:constraints}
\end{equation}%
Now, as the Euclidean distance between the boundaries (i.e., the width of the margin) follows to $\frac{2}{\|\mathbf{w}\|}$, finding the optimal separating hyperplane for the training set $\mathcal{T}$ essentially requires to minimize the length of the normal vector $\mathbf{w}$:
\begin{equation}
	\begin{split}
		\min_{\mathbf{w}\in\mathds{R}^K\backslash\{\mathbf{0}\},b\in\mathds{R}}&\|\mathbf{w}\|^2\\
		\text{s.t.}\quad\quad&(\ref{eq:constraints})\;.
	\end{split}
	\label{eq:quadratic_program}
\end{equation}%
The optimization problem (\ref{eq:quadratic_program}) is a (convex) quadratic program \cite{Boyd:Vandenberghe:04} for which a variety of solvers exists (e.g., interior point methods  \cite[p. 561]{Boyd:Vandenberghe:04}). Denoting its optimal solution as $(\mathbf{w}^{\star},b^{\star})$, we finally conclude for the desired optimal classifier (\ref{eq:classifier})
\begin{equation}
	f(\mathbf{s}) = \sign\bigl(\mathbf{w}^{\star T} \mathbf{s} + b^{\star}\bigr)\;,
	\label{eq:linclass}
\end{equation}%
where 
$\sign:\mathds{R}\to\mathds{R}$ denotes the classical signum function.
\begin{remark}\label{rem:generalization}
	It should be emphasized that learning the optimal classifier as described in this section can be easily generalized to work with training sets that cannot be separated by a hyperplane. This is typically accomplished by adding slack variables in (\ref{eq:quadratic_program}) \cite[p.\,131]{Vapnik95}. On the other hand, the method could also be modified to deal with unlabeled training sets \cite{Schoelkopf99}.
\end{remark}%
%
%
%
\section{The Wireless Channel as a Helper} \label{sec:wireless_helper}
Now, as we know how the optimal classifier looks like, the question arises how to reliably and efficiently compute it over the channel (\ref{eq:wmac}). To solve this problem, most wireless system engineers would put effort in avoiding the interference in (\ref{eq:wmac}) by employing some channel access method such as TDMA or CSMA. It means that each node gets assigned its own time slot in which it transmits its instantaneous observation to the FC that subsequently classifies the system state by evaluating (\ref{eq:linclass}). In this section, we go into the opposite direction and make explicit use of the interference by letting the nodes transmit simultaneously.

Towards this end, we adapt the transmission scheme of \cite{Goldenbaum:Stanczak:13a}, which was designed to efficiently compute real-valued functions of distributed data over the wireless channel. As shown in \cite{Goldenbaum:Boche:Stanczak:13b}, the scheme essentially allows to compute every function $f:\mathcal{S}\to\mathds{R}$ that can be represented in the form
\begin{equation}
	\label{eq:nomo}
	f(s_1,\dots,s_K) = \psi\left(\sum\nolimits_{k=1}^K\varphi_k(s_k)\right)
\end{equation}%
by a proper choice of univariate functions $\varphi_k:\mathcal{S}_k\to\mathds{R}$, $k=1,\dots,K$, and $\psi:\mathds{R}\to\mathds{R}$. Functions of this kind are known as \emph{nomographic functions} \cite{Buck:79}. 

A closer look at (\ref{eq:linclass}) reveals that the optimal classifier is nomographic with $\varphi_k(s_k)=w_k^{\star}s_k$ and $\psi(y)=\sign(y+b^{\star})$, where $w_k^{\star}$ and $s_k$ denote the $k$-th elements of $\mathbf{w}^{\star}$ and $\mathbf{s}$, respectively. In the following two subsections, we describe how this special structure permits to let the wireless channel (\ref{eq:wmac}) help in the computation. In doing so, we assume that during network initialization, the quadratic program (\ref{eq:quadratic_program}) was solved for some training set. The corresponding optimal coefficients $w_k^{\star}$ were then forwarded to the sensor nodes meaning that node $k$ knows $w_k^{\star}$ prior to network operation. On the other hand, the FC was informed about $b^{\star}$.
%
%
%
\subsection{Transmitter} \label{sec:transmitter}
In what follows, without loss of generality we focus on an arbitrary time instant $t$ and therefore drop the index for brevity. Each sensor node, say node $k$, encodes its \emph{pre-processed} observation $\varphi_k(S_k)=w_k^{\star}S_k$ as a \emph{transmit power} into a random sequence of symbols. That is, 
\begin{equation}
	X_k[m]=\frac{1}{h_k}\sqrt{g(w_k^{\star}S_k)}\exp\bigl(i\Theta_k[m]\bigr)\;,
	\label{eq:transmit_symbols}
\end{equation}%
for $m=1,\dots,M$.\footnote{Recall that $h_k\in\mathcal{C}$ was assumed to be a priori known to node $k$, $k=1,\dots,K$ (cf. Section~\ref{sec:system_model}). Thus, for $h_k\neq 0$, factor $1/h_k$ is for channel inversion. In fact, it has been shown in \cite{Goldenbaum:Stanczak:14a} that instead of $h_k$ it even suffices to know its magnitude $|h_k|$.} The purpose of function
\begin{equation*}
	g:[\varphi_{\mathrm{min}},\varphi_{\mathrm{max}}]\to[0,P_{\mathrm{max}}],g(\xi)=\frac{P_{\mathrm{max}}}{\varphi_{\mathrm{max}}-\varphi_{\mathrm{min}}}(\xi-\varphi_{\mathrm{min}})
	\label{eq:power_mapping}
\end{equation*}%
is to ensure each node satisfies the transmit power constraint $P_{\mathrm{max}}$, where
\begin{subequations}
	\begin{align}
		&\varphi_{\mathrm{min}}\coloneqq\min_{1\leq k\leq K}\min_{s\in\mathcal{S}_k}w_k^{\star}s\\
		&\varphi_{\mathrm{max}}\coloneqq\max_{1\leq k\leq K}\max_{s\in\mathcal{S}_k}w_k^{\star}s\;.%
	\end{align}%
	\label{eq:phi_range}%
\end{subequations}%
In (\ref{eq:transmit_symbols}), the random phases $\{\Theta_k[m]\}_{k,m}$ are uniformly i.i.d. over $[0,2\pi)$. Their task is to achieve a receiver-side decorrelation of the sensors' transmit signals, which simplifies the post-processing at the FC.
\begin{remark}\label{rem:power}
	Encoding the pre-processed sensor observations as a transmit power has the advantage that quantization and precise symbol and phase synchronization is not needed. For further details, the reader is referred to \cite{Goldenbaum:Stanczak:13a}.
\end{remark}%
%
%
%
\subsection{Receiver} \label{sec:receiver}
After receiving the $M$ symbols in (\ref{eq:wmac}), the FC first determines the \emph{receive energy} resulting in
\begin{equation}
	\tilde{Y}\coloneqq\sum_{m=1}^M\bigl|Y[m]\bigr|^2=M\sum_{k=1}^Kg(w_k^{\star}S_k) + \tilde{N}\;.
	\label{eq:effective_channel}
\end{equation}%
It can be easily verified that the effective noise $\tilde{N}$ has mean $\mathbb{E}\{\tilde{N}\}=M\sigma_N^2$ and variance
\setlength{\arraycolsep}{0.0em}
\begin{eqnarray}
	\hspace{-10pt}\mathbb{V}\mathrm{ar}\{\tilde{N}\}&^{}={}&2M\sum_{k=1}^K\sum_{\substack{\ell=1\\ \ell\neq k}}^K\mathbb{E}\bigl\{g(w_k^{\star}S_k)g(w_{\ell}^{\star}S_{\ell})\bigr\}\nonumber\\
	&&{+}\:2M\sigma_N^2\sum_{k=1}^K\mathbb{E}\bigl\{g(w_k^{\star}S_k)\bigr\}+M\sigma_N^4\,.
	\label{eq:variance}
\end{eqnarray}%
\setlength{\arraycolsep}{5pt}%
Except for $g$ and $\tilde{N}$, the right-hand side of (\ref{eq:effective_channel}) is already equal to the inner product $\mathbf{w}^{\star T}\mathbf{s}$. Thus, the FCs second signal post-processing step consists in applying the function
\begin{equation*}
	h:\mathds{R}\to\mathds{R}\;,\;h(\tilde{y})=\frac{\varphi_{\mathrm{max}}-\varphi_{\mathrm{min}}}{MP_{\mathrm{max}}}\tilde{y}+K\varphi_{\mathrm{min}}\;,
	\label{eq:inv_power_mapping}
\end{equation*}%
(i.e., the counterpart to $g$), which results in
\begin{equation}
	h(\tilde{Y})=\mathbf{w}^{\star T}\mathbf{s} + \frac{\varphi_{\mathrm{max}}-\varphi_{\mathrm{min}}}{P_{\mathrm{max}}M}\tilde{N}\;.
	\label{eq:inv_power_result}
\end{equation}%

Let $\alpha\coloneqq(\varphi_{\mathrm{max}}-\varphi_{\mathrm{min}})/P_{\mathrm{max}}$ and recall that $\tilde{N}$ is not zero-mean. Then, in order to account for the bias induced by $\alpha M^{-1}\tilde{N}$, we finally define the estimate of (\ref{eq:linclass}) as
\begin{equation}
	\hat{f}(\mathbf{s})=\sign\bigl(h(\tilde{Y})+b^{\star}-\alpha M^{-1}\mathbb{E}\{\tilde{N}\}\bigr)\;,
	\label{eq:f_hat}
\end{equation}%
which means that the FC finally adds $b^{\star}$ and $\alpha M^{-1}\mathbb{E}\{\tilde{N}\}$, followed by evaluating the sign of the resulting value.
%
%
%
\subsection{Proving Reliability} \label{sec:reliability}
Now, as we have introduced the transmission scheme for classifying system states by means of the wireless channel, in this subsection we prove its reliability in the sense that the corresponding probability of false classification can be made arbitrary small with increasing block length $M$.
\begin{proposition}\label{prop:reliability}
	Let $f$ be the classifier given in (\ref{eq:linclass}) and $\hat{f}$ its estimate as defined in (\ref{eq:f_hat}). Then, with the transmission scheme proposed in Sections~\ref{sec:transmitter} and \ref{sec:receiver}, any system state $\mathbf{s}\in\mathcal{S}$ can be reliably classified, in the sense of (\ref{eq:prob_condition}), over the channel (\ref{eq:wmac}).
\end{proposition}%
\begin{IEEEproof}
	Let $\mathbf{s}\in\mathcal{S}$ be arbitrary but fixed and note that 
	\begin{equation*}
		\mathbb{P}\bigl(\hat{f}(\mathbf{s})\neq f(\mathbf{s})\bigr)=\mathbb{P}\bigl(|\hat{f}(\mathbf{s})-f(\mathbf{s})|\geq 2\bigr)
		\label{eq:probs_equal}
	\end{equation*}%
	as $f(\mathbf{s})$ and $\hat{f}(\mathbf{s})$ can only take on values $\{-1,0,1\}$.\footnote{For fixed $\mathbf{s}\in\mathcal{S}$, the randomness in $\hat{f}(\mathbf{s})$ only stems from the effective channel noise $\tilde{N}$.} By virtue of Markov's inequality, it follows 
	\begin{align}
		\mathbb{P}\bigl(|\hat{f}(\mathbf{s})-f(\mathbf{s})|\geq 2\bigr)&=\mathbb{P}\bigl((\hat{f}(\mathbf{s})-f(\mathbf{s}))^2\geq 4\bigr)\nonumber\\
	&\leq\frac{1}{4}\mathbb{E}\bigl\{(\hat{f}(\mathbf{s})-f(\mathbf{s}))^2\bigr\}\;.
		\label{eq:estMarkov}
	\end{align}%
	Therefore, in order to prove that (\ref{eq:prob_condition}) applies it suffices to show that (\ref{eq:estMarkov}) vanishes with increasing block length $M$.
	
	Towards this end, substitute (\ref{eq:inv_power_result}) into (\ref{eq:f_hat}), which yields
	\begin{equation}
		\hat{f}(\mathbf{s})=\sign\left(\mathbf{w}^{\star T}\mathbf{s}+b^{\star}+\alpha M^{-1}\bigl(\tilde{N}-\mathbb{E}\{\tilde{N}\}\bigr)\right)\;.
		\label{eq:f_hat2}
	\end{equation}%
	Let $\beta\coloneqq\mathbf{w}^{\star T}\mathbf{s}+b^{\star}$ and $\Delta\coloneqq \tilde{N}-\mathbb{E}\{\tilde{N}\}$. Then, since $\alpha,M>0$, the signum function entails that
	\begin{equation}
		\hat{f}(\mathbf{s})=
		\begin{dcases*}
			-1 &if\; $\alpha\Delta/M < -\beta$\\
			0 &if\; $\alpha\Delta/M = -\beta$\\
			+1 &if\; $\alpha\Delta/M > -\beta$
		\end{dcases*}\;,
	\end{equation}%
	for all $\beta\in\mathds{R}$. Bearing this in mind, we have to show that for each of the three cases $\beta>0$, $\beta<0$, and $\beta=0$, (\ref{eq:estMarkov}) vanishes with increasing $M$. 
	
	We treat $\beta>0$ first: In this case, the true classifier provides $f(\mathbf{s})=f(\mathbf{s})^2=1$. Furthermore, for the event $\{\hat{f}(\mathbf{s})=0\}$ to occur, the point $\beta+\alpha\Delta/M$ has to lie on the separating hyperplane. Due to the distribution of $\alpha\Delta/M$, however, this event is of measure zero and therefore does not contribute to the expected value of $f(\mathbf{s})^2$. As a consequence, $\mathbb{E}\{\hat{f}(\mathbf{s})^2\}$ is equal to one and we obtain
	\begin{equation}
		\mathbb{E}\bigl\{(\hat{f}(\mathbf{s})-f(\mathbf{s}))^2\bigr\} = 2\bigl(1-\mathbb{E}\{\hat{f}(\mathbf{s})\}\bigr)\;.
		\label{eq:Case1eq1}
	\end{equation}%
	Hence, to show that (\ref{eq:estMarkov}) tends to zero relaxes to show that $\mathbb{E}\{\hat{f}(\mathbf{s})\}\to 1$ for $M\to\infty$. We have
	\begin{align*}
		\mathbb{E}\bigl\{\hat{f}(\mathbf{s})\bigr\}&=1\cdot\mathbb{P}(\alpha\Delta/M>-\beta)+(-1)\cdot\mathbb{P}(\alpha\Delta/M<-\beta)\nonumber\\
		&=2\cdot\mathbb{P}(\alpha\Delta/M>-\beta)-1\;.
		\label{eq:prob_limit}
	\end{align*}%
	Notice that for $M\to\infty$, $\alpha\Delta/M$ converges in distribution to a degenerated random variable of mean zero \cite{Goldenbaum:Stanczak:13a}.\footnote{A scalar random variable is said to be degenerated if it has finite mean and zero variance. Due to (\ref{eq:variance}) and the fact that $\alpha$ is finite, we have $\mathbb{V}\mathrm{ar}\{\alpha\Delta/M\}=\alpha^2\mathbb{V}\mathrm{ar}\{\tilde{N}\}/M^2$, which tends to zero as $M\to\infty$.} As $\beta>0$, this implies that $\mathbb{P}(\alpha\Delta/M>-\beta)$ goes to $1$ and therefore $\lim_{M\to\infty}\mathbb{E}\{\hat{f}(\mathbf{s})\}=1$ as desired. 
	
	To treat $\beta<0$, we go along similar lines as before. In this case, however, the true classifier provides $f(\mathbf{s})=-1$ so that (\ref{eq:Case1eq1}) modifies to
	\begin{equation*}
		\mathbb{E}\bigl\{(\hat{f}(\mathbf{s})-f(\mathbf{s}))^2\bigr\} = 2\bigl(1+\mathbb{E}\{\hat{f}(\mathbf{s})\}\bigr)\;.
		\label{eq:Case2eq1}
	\end{equation*}%
	For (\ref{eq:estMarkov}) to vanish, $\mathbb{E}\{\hat{f}(\mathbf{s})\}$ has to converge to $-1$ with increasing block length. It holds
	\begin{equation}
		\lim_{M\rightarrow\infty}\mathbb{E}\bigl\{\hat{f}(\mathbf{s})\bigr\}=1-2\lim_{M\rightarrow\infty}\mathbb{P}(\alpha\Delta/M<-\beta)\;.
		\label{eq:Case2eq2}
	\end{equation}%
	Now, as $\beta$ is strictly negative, the probability on the right-hand side of (\ref{eq:Case2eq2}) tends to one and thus we have $\lim_{M\to\infty}\mathbb{E}\{\hat{f}(\mathbf{s})\}=-1$ as desired. 
		
	Finally, we have to treat $\beta=0$, which means that $\mathbf{w}^{\star T}\mathbf{s}+b^{\star}$ lies on the separating hyperplane (see Fig.~\ref{fig:optimal_hyperplane}) and the true classifier provides $f(\mathbf{s})=0$. Note that from the perspective of anomaly detection, this is an unfavorable case as $\beta$ does not belong to either of the two classes ``normal'' and ``anomaly''. From the perspective of reliably computing $f$, however, we have to ensure that also in this case $\hat{f}$ provides asymptotically accurate estimates. To see that this is indeed the case notice that for $\beta=0$, (\ref{eq:f_hat2}) only depends on the sign of $\Delta$. Therefore, as $\alpha\Delta/M$ converges in distribution to a degenerate random variable of mean zero, in the limit $M\to\infty$ we have $\hat{f}(\mathbf{s})=0$ with probability one.
	
	As in each of the three cases $\mathbf{s}\in\mathcal{S}$ was chosen arbitrary, this concludes the proof.
\end{IEEEproof}%
%
%
%
\section{Numerical Examples} \label{sec:numerical_examples}
In order to evaluate and discuss the performance of the proposed classification scheme in terms of reliability and energy efficiency, in this section we present some numerical examples. As the achievable performance depends for fixed $M$ also on other system parameters such as transmit power constraint $P_{\textrm{max}}$ and noise variance $\sigma_N^2$, we declare
\begin{equation}
	\label{eq:SNR}
	\mathrm{SNR} = \frac{\frac{1}{K}\sum_k\mathbb{E}\{P_k\}}{\sigma_{N}^2}\;,
\end{equation}%
with $P_k\coloneqq g(w_k^{\star}S_k)$ the instantaneous transmit power of node $k$, as the system operating point. Notice that (\ref{eq:SNR}) is the signal-to-noise ratio of a TDMA-based scheme, averaged over nodes. 
\begin{figure}[!t]
	\centering
	\input{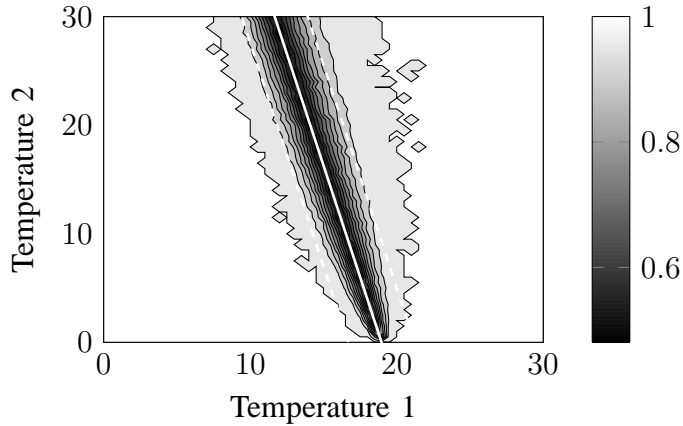} 
	\caption{Reliability map for $M=64$ and $\mathrm{SNR}=10\,\mathrm{dB}$: the solid white line represents the separating hyperplane and the dashed lines the margin, respectively.}
	\label{fig:AccOverall}
\end{figure}%
%
%
%
\subsection{Reliability} \label{sec:reliability_num}
First, we present some plots that confirm the statement of Proposition~\ref{prop:reliability}. In addition, the plots reveal how many wireless transmissions are necessary in order to achieve a certain reliability.

Without loss of generality, let the number of sensors be $K=2$, both of which are measuring a temperature in $^\circ\mathrm{C}$. Let the corresponding sensing ranges be defined as $\mathcal{S}_1=\mathcal{S}_2=[0,30]$, resulting in the state space $\mathcal{S}=[0,30]\times [0,30]$. 

We assume that each instantaneous system state, $\mathbf{s}=(s_1,s_2)^T$, belongs equally likely to either of the two classes ``normal'' and ``anomaly'', that is, $\mathbb{P}(\text{``normal''}) = \mathbb{P}(\text{``anomaly''}) = 0.5$.
The observations of class ``normal'' are distributed according to $(S_1,S_2)\sim\mathcal{N}(\boldsymbol{\mu}_1,\mathbf{\Sigma}$) and those of class ``anomaly'' as $\mathcal{N}(\boldsymbol{\mu}_2,\mathbf{\Sigma}$), respectively, with\footnote{Strictly speaking, these distributions does not provide sensor values limited to the compact state space $[0,30]\times [0,30]$. However, with the explicit choice of covariance matrix $\mathbf{\Sigma}$, values outside this space are of negligible probability.}
\begin{equation*}
	\boldsymbol{\mu}_1 = \begin{pmatrix}20\\ 20\end{pmatrix}\;,\;\boldsymbol{\mu}_2=\begin{pmatrix}10\\ 10\end{pmatrix}\;,\;\mathbf{\Sigma} = \begin{pmatrix}
1.5 & 0 \\
0 & 1.5
\end{pmatrix}\;.
\end{equation*}
In order to determine the optimal classifier (\ref{eq:linclass}), the support vector machines are trained with $L=200$ samples drawn from these distributions. For block lengths $M=2,\dots,120$ and different signal-to-noise ratios, the reliability, measured as the conditional probability 
\begin{equation}
	\mathbb{P}\left(\hat{f}(S_1,S_2)=f(S_1,S_2)\,\big|\,(S_1,S_2)=(s_1,s_2)\right)
	\label{eq:prob_error_num}
\end{equation}%
averaged over $10^4$ Monte Carlo runs each, was then evaluated for $100$ test samples.

Fig.~\ref{fig:AccOverall} shows a reliability contour plot of the $100$ test samples for $M=64$ and $\mathrm{SNR}=10\,\mathrm{dB}$. It can be seen that the estimator (\ref{eq:f_hat}) provides less reliable classifications when the system states are nearer to the separating hyperplane. In this specific example, the reliability significantly deteriorates within the margin, which is generally not the case as (\ref{eq:prob_error_num}) is mainly determined by $M$ and the SNR, whereas the margin is determined by the probability distribution of the system state.

\begin{figure}[!t]
	\centering
	\input{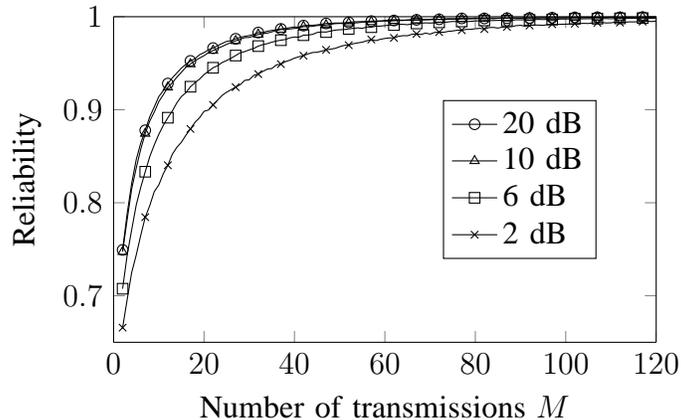} 
	\caption{Reliability as a function of the block length $M$ and different signal-to-noise ratios.}
	\label{fig:AccTestset}
\end{figure}

The trade-off between the achievable reliability, block length $M$, and the SNR is depicted in Fig.~\ref{fig:AccTestset}. As it can be seen, the reliability monotonically improves as $M$ increases. An interesting fact is that even at fairly low values of $M$ and SNR, the proposed estimator $\hat{f}$ provides results that are significantly above chance level $0.5$ (i.e., above flipping a coin).
%
%
%
%
\subsection{Energy Efficiency} \label{sec:energy_efficiency}
Now, we analyze the efficiency of the proposed classification scheme in terms of energy consumption. In doing so, a scheme that uses a simplified TDMA acts as a reference in which each sensor node gets allocated $Q\in\mathds{N}$ time slots to transmit its observation to the FC. In contrast to the preceding examples, we assume here that the network consists of $K=32$ nodes and the state space is defined to be $S=[0,30]^{32}$. Furthermore, the system states belonging to the classes ``normal'' and ``anomaly'' are distributed according to $\mathcal{N}(15\cdot\mathbf{1}_K,1.5\cdot\mathbf{I}_K)$ and $\mathcal{N}(20\cdot\mathbf{1}_K,1.5\cdot\mathbf{I}_K)$, respectively, where $\mathbf{1}_K$ denotes the length-$K$ vector of all ones and $\mathbf{I}_K$ the $K\times K$ identity matrix. For each class, 1000 samples were generated. In what follows, ``exploiting interference (EI)'' refers to the proposed scheme.

To model the energy consumption, let $P_{\mathrm{max}} = 100\,\mu\mathrm{W}$ be the peak power and $T=1\,\mathrm{ms}$ the common symbol duration. These values are in compliance with the IEEE 802.15.4 standard \cite{IEEE802154}. Then, the transmit energies follow to $E_{\text{EI},k} = MP_kT$ and $E_{\text{TDMA},k} = QP_kT$, whereas the corresponding transmit durations are given by $T_{\text{EI}} = M T$ and $T_{\text{TDMA}} = Q K T$, respectively. The overall energy consumption (per classification) is therefore given by $E_{\text{EI}} = \sum_{k=1}^K E_{\text{EI},k}$ and $E_{\text{TDMA}} = \sum_{k=1}^K E_{\text{TDMA},k}$. In order to compare the schemes in a fair manner, we set all values so that transmit durations and energy consumptions (per node) are equal (i.e., $T_{\text{EI}} = T_{\text{TDMA}} $ and $E_{\text{EI},k} = E_{\text{TDMA},k}$).

\begin{figure}[!t]
	\centering
%
%
\begin{tikzpicture}[font=\normalsize]

\begin{axis}[%
width=2.84in,
height=1.7054in,
at={(2.004in,0.746in)},
scaled ticks=false, tick label style={/pgf/number format/fixed},
scale only axis,
separate axis lines,
every outer x axis line/.append style={black},
every x tick label/.append style={font=\color{black}},
xmin=0.80,
xmax=0.99,
xlabel= {Reliability},
ylabel = {Energy Comsumption},
xtick={0.6,00.7,0.8,0.9,0.99},
every outer y axis line/.append style={black},
every y tick label/.append style={font=\color{black}},
ymin=0,
ymax=0.15,
axis background/.style={fill=white},
ylabel near ticks, yticklabel pos=left,
legend style={at={(0.03,0.65)},anchor=west,legend cell align=left,align=left,draw=black}
]
\addplot [color=black,solid, mark=triangle,mark repeat=5,thick]
  table[row sep=crcr]{%
0.706055000000001	0.0038912\\
0.778247999999999	0.0077824\\
0.824548	0.0116736\\
0.860329000000001	0.0155648\\
0.886015	0.019456\\
0.906105999999999	0.0233472\\
0.922569	0.0272384\\
0.935405999999999	0.0311296\\
0.946405	0.0350208\\
0.954491000000001	0.038912\\
0.961498000000001	0.0428032\\
0.96755	0.0466944\\
0.972831	0.0505856\\
0.976515000000001	0.0544768\\
0.980214000000001	0.058368\\
0.983108000000002	0.0622592\\
0.985522000000004	0.0661504\\
0.987728000000003	0.0700416\\
0.989529000000004	0.0739328\\
0.991035000000003	0.077824\\
0.992123000000004	0.0817152\\
0.993345000000004	0.0856064\\
0.994189000000005	0.0894976\\
0.994894000000005	0.0933888\\
0.995590000000007	0.09728\\
0.996243000000006	0.1011712\\
0.996787000000006	0.1050624\\
0.997328000000006	0.1089536\\
0.997635000000006	0.1128448\\
0.997759000000005	0.116736\\
};
\addlegendentry{TDMA};

\addplot [color=black,solid,thick]
  table[row sep=crcr]{%
0.575364	0.0002432\\
0.609236	0.0004864\\
0.632949999999999	0.0007296\\
0.654781000000001	0.0009728\\
0.67181	0.001216\\
0.687206000000001	0.0014592\\
0.701531	0.0017024\\
0.714559	0.0019456\\
0.725456999999999	0.0021888\\
0.736097999999999	0.002432\\
0.745860999999999	0.0026752\\
0.755774999999999	0.0029184\\
0.764228	0.0031616\\
0.772655	0.0034048\\
0.779815000000001	0.003648\\
0.787917999999999	0.0038912\\
0.794318000000001	0.0041344\\
0.800708	0.0043776\\
0.807671	0.0046208\\
0.814134999999999	0.004864\\
0.819292999999999	0.0051072\\
0.825064	0.0053504\\
0.829914999999999	0.0055936\\
0.83481	0.0058368\\
0.840083999999999	0.00608\\
0.844348	0.0063232\\
0.848646999999999	0.0065664\\
0.853514999999999	0.0068096\\
0.857395	0.0070528\\
0.861281	0.007296\\
0.865251000000001	0.0075392\\
0.869281	0.0077824\\
0.872480999999999	0.0080256\\
0.876394000000001	0.0082688\\
0.879181999999999	0.008512\\
0.882337	0.0087552\\
0.885507999999999	0.0089984\\
0.889121999999998	0.0092416\\
0.891546999999998	0.0094848\\
0.894818999999999	0.009728\\
0.897736	0.0099712\\
0.899590000000001	0.0102144\\
0.902548	0.0104576\\
0.905009999999999	0.0107008\\
0.907278000000002	0.010944\\
0.910098000000001	0.0111872\\
0.912122999999999	0.0114304\\
0.914258999999999	0.0116736\\
0.916691	0.0119168\\
0.918567999999999	0.01216\\
0.920116	0.0124032\\
0.922466	0.0126464\\
0.924036	0.0128896\\
0.926244	0.0131328\\
0.927488999999999	0.013376\\
0.92967	0.0136192\\
0.930888000000001	0.0138624\\
0.932923000000002	0.0141056\\
0.933930000000001	0.0143488\\
0.935796999999999	0.014592\\
0.937208000000001	0.0148352\\
0.938586000000001	0.0150784\\
0.940137000000002	0.0153216\\
0.941847	0.0155648\\
0.943278	0.015808\\
0.944481000000002	0.0160512\\
0.945694999999999	0.0162944\\
0.946373	0.0165376\\
0.947955999999999	0.0167808\\
0.949629000000001	0.017024\\
0.950451	0.0172672\\
0.951617000000001	0.0175104\\
0.952688000000002	0.0177536\\
0.953307	0.0179968\\
0.954658	0.01824\\
0.955622	0.0184832\\
0.95664	0.0187264\\
0.957315	0.0189696\\
0.958591000000002	0.0192128\\
0.959533000000001	0.019456\\
0.960329000000001	0.0196992\\
0.961136000000001	0.0199424\\
0.962032	0.0201856\\
0.963071000000002	0.0204288\\
0.963599000000001	0.020672\\
0.964527000000001	0.0209152\\
0.965285000000003	0.0211584\\
0.966111999999999	0.0214016\\
0.966770000000002	0.0216448\\
0.967349000000002	0.021888\\
0.968259	0.0221312\\
0.968502000000002	0.0223744\\
0.969500000000002	0.0226176\\
0.970150000000002	0.0228608\\
0.970408000000002	0.023104\\
0.971070000000003	0.0233472\\
0.971871000000003	0.0235904\\
0.972562000000002	0.0238336\\
0.973347000000002	0.0240768\\
0.973824000000002	0.02432\\
0.974364000000002	0.0245632\\
0.974838000000003	0.0248064\\
0.975384000000002	0.0250496\\
0.975894000000003	0.0252928\\
0.976426000000003	0.025536\\
0.976582000000003	0.0257792\\
0.977236000000002	0.0260224\\
0.977661000000003	0.0262656\\
0.978065000000003	0.0265088\\
0.978600000000002	0.026752\\
0.978889000000003	0.0269952\\
0.979369000000003	0.0272384\\
0.979624000000003	0.0274816\\
0.980122000000003	0.0277248\\
0.980595000000002	0.027968\\
0.981072000000003	0.0282112\\
0.9815	0.0284544\\
0.9822	0.0286976\\
0.9826	0.0289408\\
0.983	0.029184\\
0.9834	0.0294272\\
0.983499000000004	0.0296704\\
0.983630000000003	0.0299136\\
0.984119000000004	0.0301568\\
0.984524000000003	0.0304\\
0.984848000000002	0.0306432\\
0.985265000000003	0.0308864\\
0.985281000000004	0.0311296\\
0.985631000000003	0.0313728\\
0.986276000000002	0.031616\\
0.986538000000005	0.0318592\\
0.986724000000003	0.0321024\\
0.987052000000003	0.0323456\\
0.987027000000003	0.0325888\\
0.987434000000003	0.032832\\
0.987761000000003	0.0330752\\
0.987747000000004	0.0333184\\
0.988126000000005	0.0335616\\
0.988389000000004	0.0338048\\
0.988740000000003	0.034048\\
0.988989000000004	0.0342912\\
0.988965000000003	0.0345344\\
0.989566000000003	0.0347776\\
0.989564000000004	0.0350208\\
0.989674000000004	0.035264\\
0.989846000000005	0.0355072\\
0.989956000000004	0.0357504\\
0.990305000000003	0.0359936\\
};
\addlegendentry{Exploit. Int.};

	\node[align=left, above] at (axis cs:0.915,0.05){$E_{\text{TDMA}} = 2.11\cdot E_{\text{EI}}$};
  
  \draw[<->, style=thick, color=gray, thick, dashed](axis cs: 0.98,0.0283)--(axis cs: 0.98,0.0575);
  \draw[->, style=thick,color=gray, thick](axis cs: 0.92,0.05)--(axis cs: 0.978,0.042);

\end{axis}
\end{tikzpicture}%
	\caption{Exploiting Interference (EI) vs. TDMA: energy consumption over reliability for $K=32$ and $\mathrm{SNR}=1\,\mathrm{dB}$. At a reliability of $98\,\%$, exploiting interference requires $53\,\%$ less energy than avoiding it.}
	\label{fig:ComacVsTDMAenergy}
\end{figure}
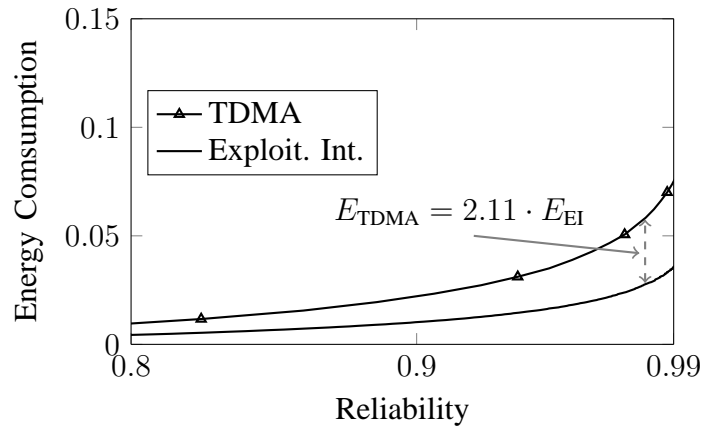

As can be seen from Fig.~\ref{fig:ComacVsTDMAenergy}, the scheme proposed in this paper is much more energy efficient as the TDMA-base scheme. For instance, to achieve a reliability of $98\,\%$ exploiting interference only requires $47\,\%$ of the energy required with TDMA, due to the  smaller number of transmissions required. In simple terms, the wireless channel helps to significantly improve the reliability as well as the energy efficiency.
%
%
%
\section{Conclusions} \label{sec:conclusions}
In this paper, the problem of reliably and efficiently classifying a system state for anomaly detection was considered. The problem occurs in many relevant areas such speech and image recognition, condition monitoring, and fault diagnosis. To solve the problem, a novel classification scheme was proposed that is based on a set of wireless sensor nodes that monitor the system and simultaneously transmit their observations to a fusion center. In contrast to similar schemes known from the literature, the proposed scheme considers the wireless channel as helper. This means that the optimal classifier is essentially computed in the air by exploiting interference rather than avoiding it. The optimal classifier was learned from the sensor data based on the celebrated support vector machines model. Given this classifier, a corresponding estimator was proposed and shown to be able to make the probability of false classification arbitrary small. Finally, some numerical examples were presented that indicate that the proposed classification scheme is much more energy efficient as a scheme based on time-division multiple access. As a consequence, considering the wireless channel as a helper when transmitting the sensor observations to the fusion center may significantly increases the reliability as well as the energy efficiency. 
%
%
%

%
%
\end{document}